\documentclass{aastex}
\usepackage{emulateapj5,apjfonts}

\newcommand{\rmn}[1]{\mathrm {#1}}

\shorttitle{NONLINEAR THERMAL INSTABILITY IN THREE DIMENSIONS}
\shortauthors{KRITSUK \& NORMAN}

\begin{document}
\def\lsim{~\raise0.3ex\hbox{$<$}\kern-0.75em{\lower0.65ex\hbox{$\sim$}}~}
\def\gsim{~\raise0.3ex\hbox{$>$}\kern-0.75em{\lower0.65ex\hbox{$\sim$}}~}

\journalinfo{The Astrophysical Journal, 569:L00-L00, 2002 April 20}

\submitted{Received 2001 December 18; accepted 2002 March 20; published 2002
March 00}
\title{Thermal Instability--Induced Interstellar Turbulence}

\author{Alexei G. Kritsuk and Michael L. Norman}
\affil{Department of Physics and Center for Astrophysics and Space Sciences, 
University of California at San Diego,\\
9500 Gilman Drive, La Jolla, CA 92093-0424; 
akritsuk@ucsd.edu, mnorman@cosmos.ucsd.edu}

\begin{abstract}
We study the dynamics of phase transitions in the interstellar medium
by means of three-dimensional hydrodynamic numerical simulations.
We use a realistic cooling function and generic nonequilibrium
initial conditions to follow the formation history of a multiphase medium
in detail in the absence of gravity.
We outline a number of qualitatively distinct stages of this process, 
including a linear isobaric evolution, transition to an isochoric regime,
formation of filaments and voids (also known as ``thermal'' pancakes), 
the development and decay of supersonic turbulence, an approach to pressure 
equilibrium, and final relaxation of the multiphase medium. 
We find that 1\%-2\% of the initial thermal energy is converted into gas 
motions in one cooling time.
The velocity field then randomizes into turbulence that decays on a dynamical 
timescale $E_k\propto t^{-\alpha}$, $1\lsim\alpha\lsim2$.
While not all initial conditions yield a stable
two-phase medium, we examine such a case in detail. 
We find that the two phases are well mixed with the cold clouds
possessing a fine-grained structure near our numerical resolution limit. 
The amount of gas in the intermediate unstable phase roughly
tracks the {\em rms} turbulent Mach number, peaking at 25\% when 
${\cal{M}}_{rms}\sim8$,
decreasing to 11\% when ${\cal{M}}_{rms}\sim0.4$.
\end{abstract}


\section{Introduction}
Thermal instability (TI) has many implications in astrophysics (e.g., a clumpy
interstellar medium [ISM], stellar atmospheres, star formation, globular 
cluster and galaxy formation, etc.; see Meerson 1996 
for a recent review). 
The instability may be driven by radiative cooling of optically thin
gas (radiation-driven TI) or by exothermic nuclear reactions 
\citep{schwarzschild.65}.

Linear stability theory for a medium with volumetric sources and sinks of
energy in thermal equilibrium was developed 
by \citet{field65}, who identified three unstable modes: 
the {\em isobaric} mode (the pressure-driven formation of condensations 
not involving gravitation) and the two {\em isentropic} modes (the 
overstability of acoustic waves propagating in opposite directions).
Hunter (1970, 1971) extended these results to an arbitrary
nonstationary background flow, showing that cooling-dominated media are
potentially more unstable than those in equilibrium, while heating provides
stabilization.
The most common applications of thermal instability to the ISM and star 
formation deal with the isobaric mode that was employed to explain the 
observed multiphase structure of the ISM 
\citep{pikel'ner68,field..69,mckee.77,mckee90,wolfire....95}.

Analysis of infinitesimal perturbations gives two characteristic
length scales for the isobaric condensation mode: 
(1) a cooling scale $\lambda_{\rmn{p}}=c/\omega_{\rmn{p}}$ 
(where $c$ is the adiabatic sound speed and $\omega_{\rmn{p}}$ is the 
growth rate) and
(2) a critical scale $\lambda_{\kappa}=c\sqrt{t_{d}/\omega_{\rmn{p}}}$
(where $t_{d}$ is the characteristic thermal diffusion time).
These two length scales define short-, intermediate-, and long-wavelength 
limits \citep{meerson96,kovalenko.99}.
In the short-wavelength limit, small isobaric perturbations are inhibited by 
heat conduction, so that $\omega_{\rmn{p}}<0$ for $\lambda<\lambda_{\kappa}$.
In the long-wavelength limit, the perturbations cannot grow isobarically
because of the finite sound speed effects, and thus 
$\omega_{\rmn{p}}\rightarrow0$ for 
$\lambda/\lambda_{\rmn{p}}\rightarrow\infty$.
This is only true if the gas is {\em isochorically} stable 
\citep{parker53,field65,shchekinov78}, 
otherwise the growth rate remains finite:
$\omega_{\rmn{p}}\rightarrow\omega_{\rmn{\rho}}>0$ for 
$\lambda/\lambda_{\rmn{p}}\rightarrow\infty$, but only
large-scale temperature perturbations are growing, thus resulting in
pressure variations and the formation of shock waves.
The growth rates and characteristic scales depend on the heating and 
cooling properties of a given medium.
Under the ISM conditions, if one assumes thermal equilibrium
(i.e., an exact balance between cooling and heating), isochoric instability 
manifests itself only at relatively high temperatures, $T \gsim 10^5$~K.
However,  in cooling-dominated regimes, it can develop at temperatures
as  low as $10^3$~K.

\begin{figure*}
\epsscale{2.0}
\plotone{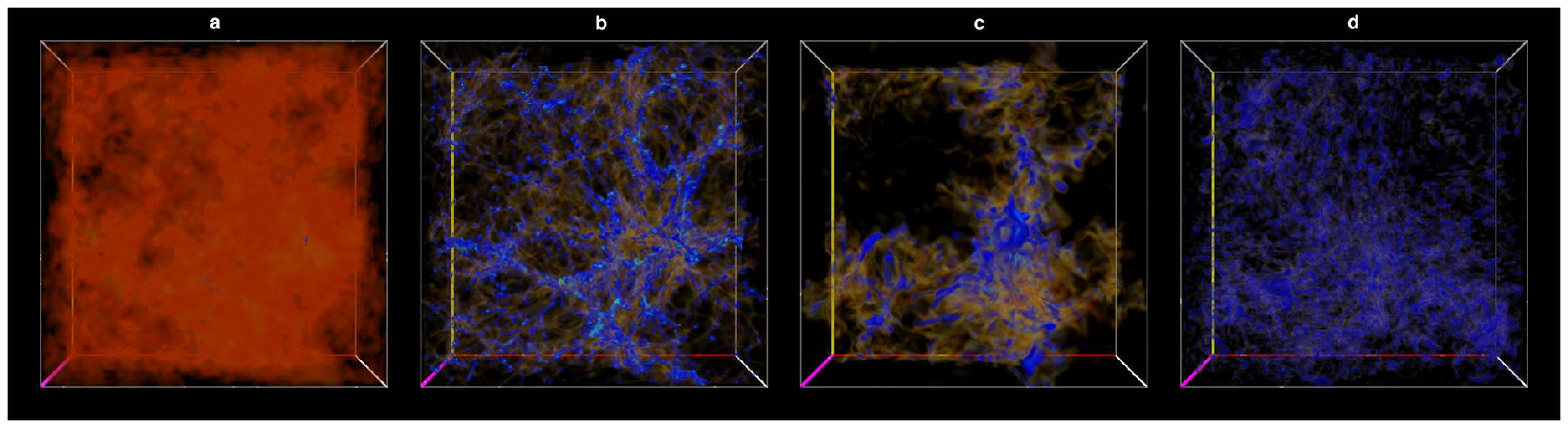}
\caption{Snapshots of the gas density field (perspective volume rendering):
(a) First condensation at $t=0.07$~Myr,
(b) thermal pancakes at $t=0.1$~Myr,
(c) collapse and turbulization of cellular structure at $t=0.17$~Myr,
(d) two-phase medium at $t=1.5$~Myr
(5~pc box, $256^3$ grid points).
The log density color coding is as follows: 
The dense blobs at the intersections of the filaments, 
$\rho>10^{-22}$~g cm$^{-3}$, are light blue;
the stable cold phase, $\rho\in[10^{-23}, 10^{-22}]$~g cm$^{-3}$, is blue;
the unstable density regime, 
$\rho\in[10^{-23.7}, 10^{-23.0}]$~g cm$^{-3}$, is yellow to brown;
and the low-density gas, including stable warm phase 
($\rho<10^{-23.7}$~g cm$^{-3}$), is a transparent red.
The figure is also available as an mpeg animation in the electronic 
edition of the Astrophysical Journal.
\label{fig1}} 
\end{figure*}

A specific feature of TI in the ISM is a large 
($\gsim1$ dex) gap in gas densities between the two stable phases.
The density range of interest in a galaxy formation context is even larger.
This implies the importance of nonlinear effects in the dynamics of phase 
transitions.
Nonlinearity brings into play nonequilibrium effects.
Already weakly nonlinear development of condensations in an initially 
homogeneous gas in thermal equilibrium drives the system away from
equilibrium.
The mean pressure drops since $\bar{\rho^2} > \bar{\rho}^2$ 
and cooling overcomes heating globally \citep{kritsuk85}.
Later, on a timescale of $\sim\omega_{\rmn{p}}^{-1}$, as condensations 
get denser and cooler, the isobaric condition $\lambda\ll\lambda_{\rmn{p}}$
becomes violated locally within them, so the system departs from pressure 
equilibrium.
These effects are essential for the isobaric mode of TI in the ISM and star 
formation contexts. 
Therefore, analytical nonlinear solutions to ``isobarically'' reduced
TI equations are insufficient to describe the radiative stage of the 
phase transition ($t\sim\omega_{\rmn{p}}^{-1}$).

During this strongly nonlinear stage {\em large-scale}
condensations form in such a way that gas moves almost inertially and
its kinetic energy dominates thermal energy ($p\ll \rho v^2$).
Accordingly, the gas velocities in these condensations are of the order of 
the sound speed in the unperturbed state.
The situation here is directly analogous to the long-wave gravitational 
instability, so that results concerning the origin of cellular structure
and \citet{zel'dovich70} ``pancakes'' can be entirely carried over to 
the case of long-wave TI \citep{meerson.87}.
\citet{sasorov88} gave an elegant proof that qualitatively the same result 
applies to {\em small-scale} TIs; i.e., the onset of 
TI produces voids and highly flattened condensations along 
certain two-dimensional surfaces.
These are also called ``thermal'' pancakes \citep{meerson96}.
The formation of filaments was simultaneously noticed in two-dimensional 
numerical simulations of TI in the solar transition region 
\citep{dahlburg....87,karpen..88}.
Ever since, thermal pancakes are being rediscovered both analytically and in
numerical simulations (e.g., Lynden-Bell \& Tout 2001).

Thermal pancakes are transient. 
However, what happens next, before the evolution turns to a conductive 
relaxation stage \citep{meerson96}, until recently has remained the
``terra incognito'' of TI theory. 
The problem of ``postradiative'' mechanical relaxation toward a static 
multiphase medium requires a solution for the full set of hydrodynamic 
equations that can only be obtained numerically.
One-dimensional simulations pioneered by \citet{goldsmith70} demonstrated
that TI develops large motions in the ISM (see also Hennebelle \& P{\'e}rault
1999 for an example of how large motions can trigger TI). 
For some time, progress in this direction was precluded by numerical 
difficulties in modeling convergent cooling flows with high Mach numbers and 
high-density contrasts (e.g., V{\'a}zquez-Semadeni, Gazol, \& Scalo  2000).
Recent multidimensional numerical simulations of the ISM evolution in 
disklike galaxies include effects of gravity, differential rotation, star 
formation and supernova feedback (de Avillez 2000; 
V{\'a}zquez-Semadeni et al. 2000;
Wada \& Norman 2001; Wada 2001).
However, it is hard to determine the role of TI in shaping the ISM structures
found in these models, partly because the simulations still do not resolve 
length scales important for TI and partly because of the additional physical 
effects.

The purpose of this Letter is to report on results of three-dimensional
numerical simulations of classical TI that fill the gap in theory, exploring
in detail the late radiative stage and postradiative relaxation toward a 
multiphase medium.
Our major result is that formation of thermal pancakes induces turbulence in
the ISM that serves as a nonlinear saturation mechanism for TI.
As a consequence of a turbulent cascade, 
(1) information about initial perturbations is lost, including the 
imprints of heat conduction in the density power spectrum during the
linear stage, and
(2) turbulent diffusion becomes the dominant transport 
mechanism during the postradiative relaxation stage.

\begin{figure*}
\epsscale{1.8}
\vspace{-1.8cm}
\plottwo{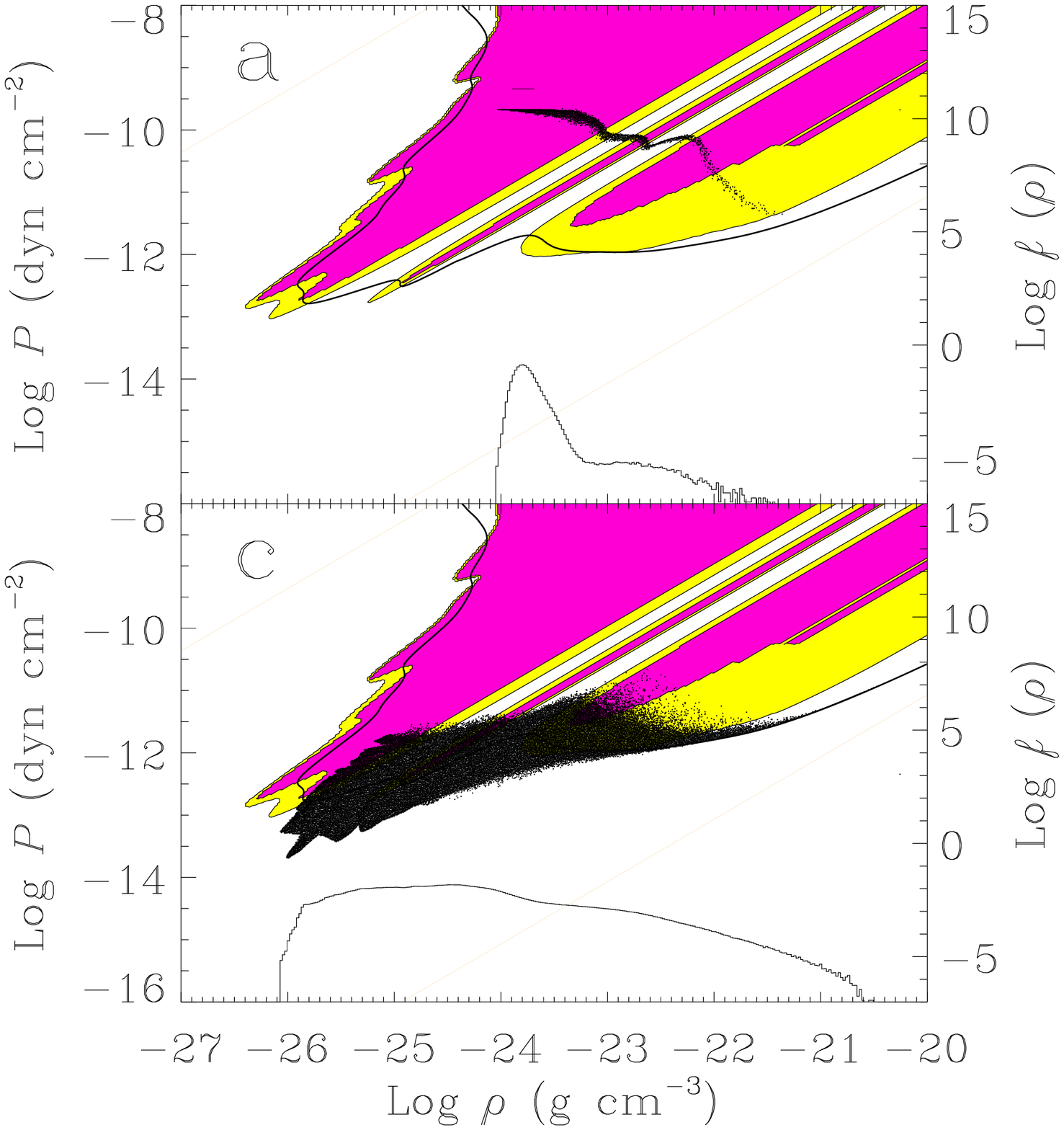}{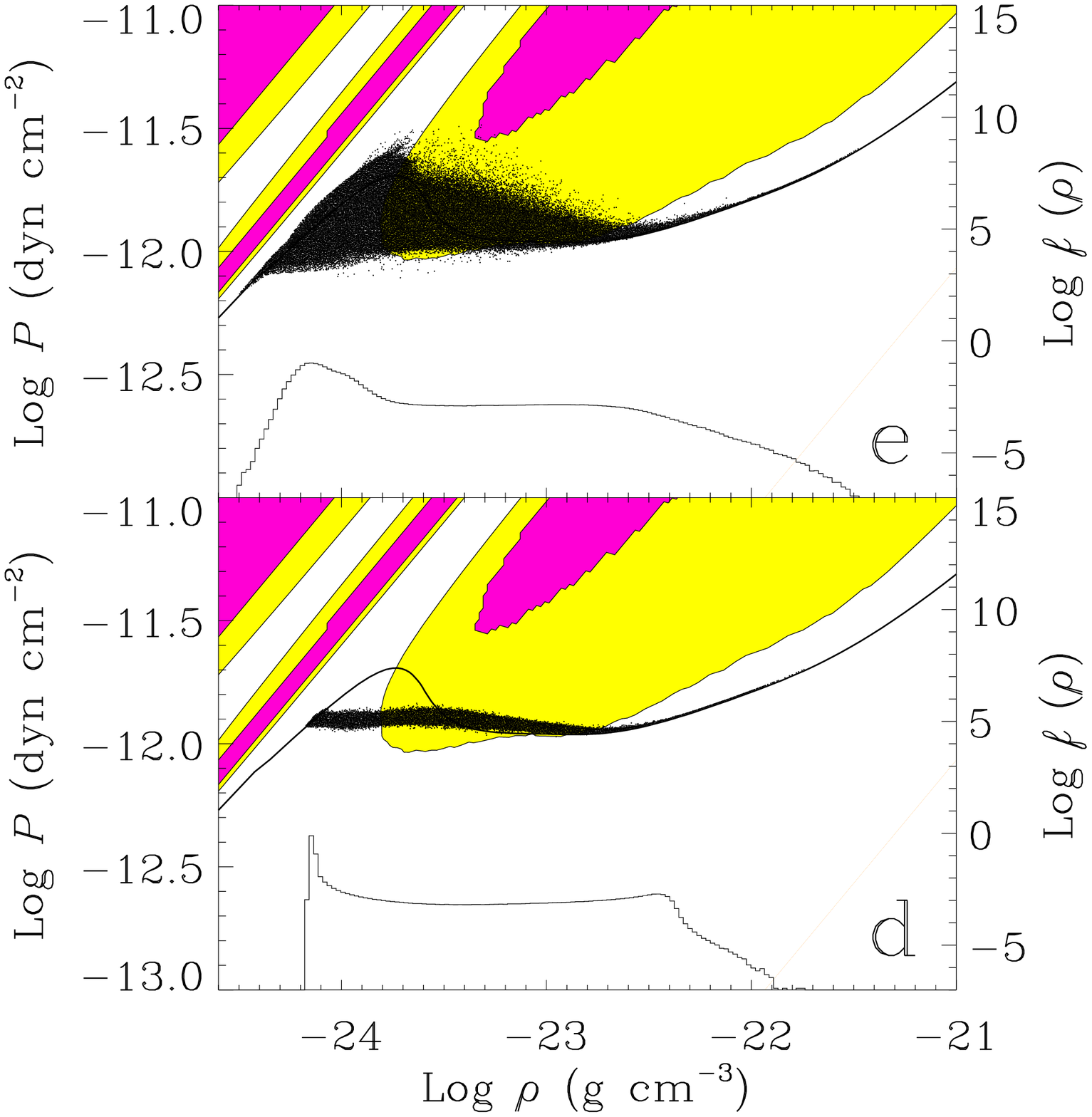}
\caption{Snapshots of phase diagrams (timing and labels correspond to those 
in Fig. \ref{fig1}): (a) $t=0.07$~Myr, (c) $t=0.17$~Myr, (e) $t=0.5$~Myr,
(d) $t=1.5$~Myr.
The black dots show scatter plots of pressure vs. density.
The ``dash'' at $P=4.55\times10^{-10}$~dyn~cm$^{-2}$ in (a) shows 
the isobaric initial conditions.
Background yellow-filled contours specify the part of the phase plane 
where isobaric mode is unstable; 
overlaid magenta contours are the regions of isochoric instability.
The thick solid line shows thermal equilibrium curve.
Density PDFs are plotted at the bottom of each panel (see scale to the right).
\label{fig2}} 
\end{figure*}

\section{Simulations}
We solve the equations of ideal gasdynamics (eqs. [6]-[9] in 
Field 1965) in a cubic domain with periodic boundary conditions, 
assuming zero conductivity and no gravity.
The generalized cooling function ${\cal L}$ contains cooling and heating terms,
${\cal L}= \rho\Lambda(T,Z) - \Gamma$.
The temperature dependence of radiative cooling $\Lambda(T,Z)$ for 
$T\in[10, 10^8]$~K is provided by M. Spaans (2000, private communication;
see also Wada \& Norman 2001). 
The gas metallicity $Z$ is assumed to be solar in most of the models, 
but we also considered a case with a subsolar value of $0.1Z_{\odot}$.
For simplicity we assume that the heating rate $\Gamma\equiv const$, 
and we define it in units of the cooling rate in an initial unperturbed state:
$\Gamma=Q\rho_0\Lambda(T_0,Z)$, where $Q\in\{0, 0.01, 0.1, 0.5, 1 \}$.
The initial unperturbed state is uniform with gas density
$\rho_0\in\{0.167, 1.67\}\times10^{-24}$~g~cm$^{-3}$ and gas 
temperature $T_0\in\{7\:000,\; 2\times10^6\}$~K.
The box size $L\in\{5, 100, 500\}$~pc was chosen such that at least the 
cooling scale $\lambda_{\rmn{p},0}$ for initial state is well resolved 
on a discrete grid of $128^3$ or $256^3$ cells.
In most cases, we initially impose isobaric density perturbations with a 
three-dimensional power spectrum index of $-3$, 
an {\em rms} deviation of $\varepsilon=0.05$, and a high-wavenumber cutoff
$k_{max}\in\{8,32\}$,\footnote{These are normalized values; $k_{max}=32$ for
a grid $256^3$ implies that the cutoff scale spans eight grid zones. 
This is slightly below the dissipation limit of our numerical scheme.} 
assuming zero initial velocities.
A few cases were started up with random velocity perturbations, varying the
amplitude and assuming uniform initial density.

The simulations were performed using two hydrosolvers implemented in
the Enzo code developed by \citet{bryan.00}: one is a direct 
Eulerian version of the PPM scheme \citep{colella.84}, and the other is the 
finite difference Eulerian method originally implemented in ZEUS 
\citep{stone.92}.

\section{Results}
Here we present results for our fiducial model computed on a 
``moderate''-resolution grid of $256^3$ cells assuming $L=5$~pc, 
$\rho_0=1.67\times10^{-24}$~g~cm$^{-3}$, $T_0=2\times10^6$~K, $Q=0.1$, solar
metallicity, and isobaric perturbations with $\varepsilon=0.05$, $k_{max}=32$
(Fig. \ref{fig1}).
With these initial conditions, the linear isobaric mode develops on a 
timescale $\omega_{\rmn{p},0}^{-1}\approx0.3$~Myr, 
the cooling length $\lambda_{\rmn{p},0}\approx70$~pc, and the critical
length for heat conduction $\lambda_{\kappa,0}\approx8$~pc.
Thus, our initial perturbations formally drop into a short-wavelength 
limit.\footnote{Since initial perturbations are not infinitesimal, 
$\frac{\delta\rho}{\rho}|_{max,0}=0.23$, TI reaches the strongly nonlinear 
regime by 0.07~Myr. 
For such perturbations, both $\lambda_{\rmn{p},0}$ and 
$\lambda_{\kappa,0}$ are effectively lower than their linear estimates.}
Assumed cooling and heating rates and the value of the initial gas density 
imply the existence 
of attracting bistable thermal equilibrium, with physical conditions close 
to those in the ISM (Fig. \ref{fig2}), although the pressure range for 
bistability is somewhat narrower than in the ``standard'' model of 
\citet{wolfire....95}.

\begin{figure*}
\epsscale{1.8}
\vspace{-2.2cm}
\plottwo{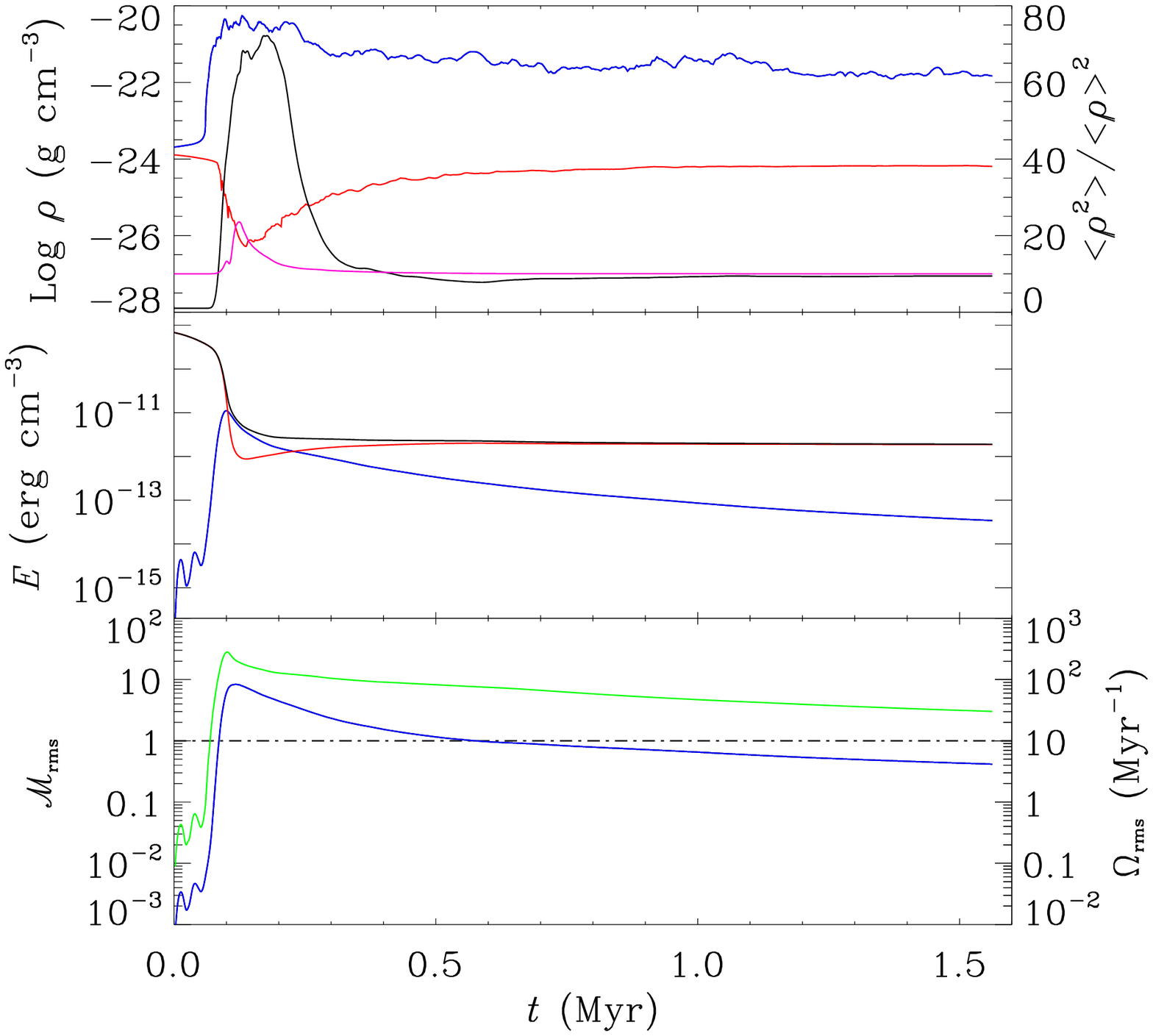}{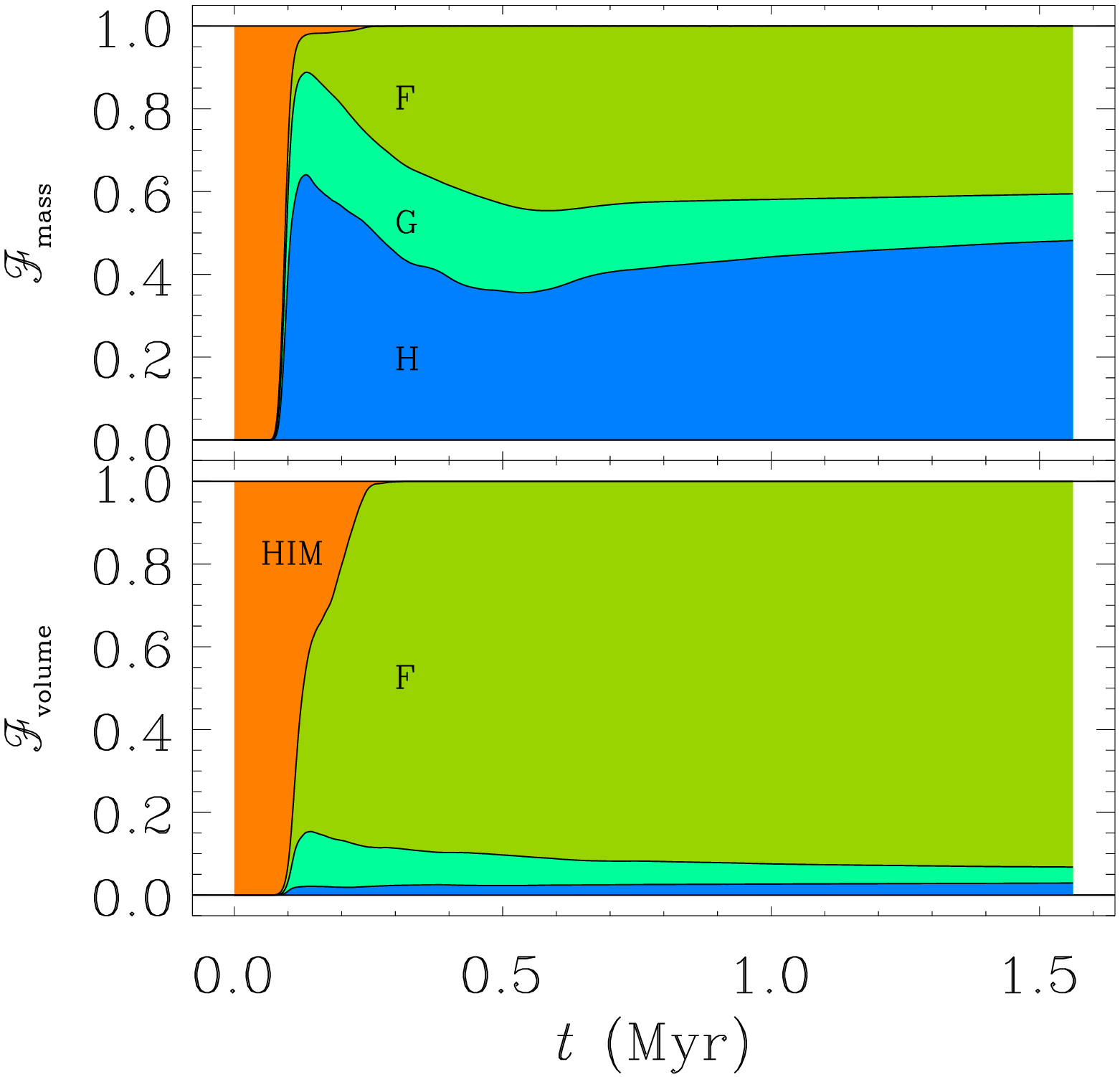}
\caption{Time evolution of global variables. {\em Top left panel}: 
$\rho_{max}$ ({\em blue}),
$\rho_{min}$ ({\em red}), $<\rho^2>/<\rho>^2$ ({\em black}), 
$10<p^2>/<p>^2$ ({\em magenta},
scale is at the right).
{\em Middle left panel}: Total energy ({\em black}), 
thermal energy ({\em red}), kinetic
energy ({\em blue}).
{\em Bottom left panel}: Mass-weighted {\em rms} Mach number ({\em blue}), 
{\em rms} enstrophy ({\em green}).
{\em Top right panel}: Mass fractions of different thermal phases:
hot (labeled HIM, $T>19,000$~K)
warm stable (labeled F, $T\in[8\:000, 19,000]$~K), 
intermediate unstable (labeled G, $T\in[600, 8\:000]$~K), 
and cold stable (labeled H, $T<600$~K).
{\em Bottom right panel}: Same as above, but for volume fractions. 
\label{fig3}} 
\end{figure*}

With our setup, heating does not compensate cooling in the initial state; 
thus, TI sets in on a time-dependent background state.
As the gas cools, the instability channels a part of its thermal energy 
into kinetic energy of converging flows that create condensations
and evacuate gas from underdense voids (Fig.\ref{fig3}).
Growing density variance accelerates cooling efficiency, so the gas as a 
whole cools much faster than a homogeneous medium with the same initial setup.
Since initial perturbations span a range of linear scales, the epoch of the
thermal pancakes' formation lasts from $\sim0.07$~Myr, when the first cold blob
forms, to $\sim0.1$~Myr, when the mean kinetic energy and then pressure 
variance reach their maximal values (see Figs. \ref{fig1}a, \ref{fig1}b,
\ref{fig2}a, and~\ref{fig3}).

The pancakes themselves exhibit a rather complex inner structure.
As the isobaric compression gives way to isochoric cooling
\citep{kritsuk90,burkert.00}, 
accretion shocks develop within the cooling condensations at temperatures 
corresponding to two strips of stability in the phase plane (Fig. \ref{fig2}a).
Dense gas in the cores of condensations cools further as it contracts in a 
regime similar to the explosive condensation described in \citet{meerson.87}; 
see Figure \ref{fig2}a.

Due to asymmetries in the initial conditions, the dense cores gain nonzero 
momentum that drives a bottom-up collapse of the hierarchical cellular 
structure composed of thermal pancakes.
As a result, a single large void forms in the periodic box, and $\rho_{min}$
attains its global minimum by $t=0.13$~Myr (Figs. \ref{fig1}c and
\ref{fig3}).
At this point, the gas density variance approaches its maximum, 
and the density spans a range of 5.5 dex.
The highest density gas quickly relaxes to thermal equilibrium. 
However, the dense cold blobs reexpand slightly until they reestablish pressure
balance with the less dense environment, so the mass fraction of the gas in
cold stable phase ``H'' decreases\footnote{We follow the notation of 
Field et al. (1969).} (see Fig.~\ref{fig3}).

By $t\approx0.15$~Myr, the information about the details of the 
initial perturbations is lost, and highly compressible supersonic turbulence 
with an {\em rms} Mach 
number of about 10 is fully developed in the computational domain 
(Fig. \ref{fig3}). 
The velocity power spectrum fills in at small scales as the turbulent cascade 
settles and then only slightly evolves on a time scale of $\sim1$~Myr as 
turbulence decays [$E_{k}\approx9.0\times10^{-13} 
(t/0.3\;\rmn{Myr})^{-2}$~ergs~cm$^{-3}$ for
$t\in(0.3, 1.7)$~Myr, cf. Mac Low, Klessen, \& Burkert 1998].
The density probability distribution function (PDF) exhibits a power-law
excess at high densities, where the ``effective'' equation of state is 
soft\footnote{For 
$\Lambda(T)\propto T^{\alpha}$ and $\Gamma=const$, the equilibrium pressure 
$p_{eq}\propto\rho_{eq}^{1-\frac{1}{\alpha}}$, 
and since $\alpha>1$ at low temperatures, 
the effective adiabatic index $\gamma_{eq}<1$.}.
Low-density gas is nearly adiabatic ($\gamma =\frac{5}{3}$); therefore, 
the PDF also has a power-law regime at low $\rho$ (Figs. \ref{fig2}c and
\ref{fig2}e).
The power-law regimes appear in response to deviations from a special 
{\em isothermal} case in which {\em lognormal} PDFs are produced due to the 
fact that the local Mach number is independent of the density 
\citep{passot.98}.

Relaxation timescales toward thermal and dynamical equilibria depend on
local physical conditions.
Cold dense gas quickly settles to a thermal equilibrium, but its dynamical
relaxation proceeds quite slowly (by $t=1.5$~Myr, its kinetic energy dominates 
the thermal energy by up to a factor of 3 in the most dense blobs).
Warm tenuous gas, instead, undergoes a fast transition to a subsonic regime
with a quite uniform density distribution (see PDFs in Figs. \ref{fig2}e and 
\ref{fig2}d), but it takes longer to establish thermal equilibrium.
As soon as the system relaxes to a bistable state with pressure close to
the allowed minimum,\footnote{The pressure hovers near $P_{min}$ because of
the shape of the thermal equilibrium curve, which is in turn
controlled by our assumed heating and cooling functions.} 
the density PDF shows a bimodal distribution, phase 
fractions by mass saturate at $\sim42\%$ for the warm phase, and $\sim44\%$ 
for the cold phase, and $\sim14\%$ of the mass falls in the unstable 
temperature regime.
The substantial amount of gas in the unstable regime can be explained as 
being due to turbulent diffusion.
Vorticity is generated by baroclinic instabilities and in
shocks, which develop during the radiative stage at the interface of 
growing condensations (Dahlburg et al. 1987; Karpen et al. 1988).
The values we found for the unstable gas fraction are somewhat lower than
those obtained in two-dimensional models by \citet{gazol...01}.
To see whether turbulence was at least partially driven by mass exchange 
between phases, we zeroed the velocity field at $t\sim1.5$~Myr.
Residual pressure variations regenerated the turbulence, but at a much lower
level.
The gas mass fraction in the unstable regime ``G'' 
shrunk from 14\% to only 2\%;
the remainder is incorporated into the cold stable phase H.
We conclude that TI-induced turbulence is purely decaying in our simulations.

Since at least some part of the gas does not stay in thermal equilibrium
(Fig. \ref{fig2}d), the exact fraction of thermally unstable gas should be
determined by Hunter's (1970) TI criterion.
A rough estimate, however, could be obtained using density and temperature 
limits for the stability of thermal equilibrium, which are used in
Figs \ref{fig1} and \ref{fig3}, respectively.
Note that such estimates are robust for only a medium in a quasi-isobaric 
state.
One has to be cautious obtaining temperature estimates for the unstable gas 
based on the assumption of thermal equilibrium (cf. Heiles 2001a).

Only because we were able to resolve the length scales that were smaller 
than the cooling scale $\lambda_{\rmn{p}}$ in the vicinity of
the bistable regime did the gas in the box relax to an isobaric 
state.\footnote{
It was the necessity to resolve isobaric mode in the vicinity of bistable
thermal equilibrium that forced us to choose the box size $L=5$~pc.}
Seed perturbations in the corresponding range of linear scales are always
available to feed the instability, provided the turbulent cascade exists.
The resulting two-phase medium is dynamic. 
Turbulent velocities do not correlate with density, and dense clouds 
appear to be shapeless random aggregations of cold Lagrangian gas parcels,
forming a ``fractal'' substrate (Fig.~\ref{fig1}d). 

\section{Discussion}

Our fiducial case was constructed to produce a stable two-phase medium because
of its relevance to the Galactic ISM. We are interested in TI over a wide 
range of conditions as might be found in the ISM of
high-redshift protogalaxies. 
We have simulated other cases with different parameter choices that do not
produce stable two-phase media. However, we find they all develop turbulence 
in the nonlinear radiative stage of TI. Here we briefly discuss how the 
turbulence and asymptotic
phase structure depend on initial conditions, deferring a more complete
discussion to a future paper (A. Kritsuk \& M. Norman 2002, in preparation). 

The level of induced turbulence is determined by the efficiency of
conversion of the initial thermal gas energy into kinetic energy of
turbulent flow by nonlinear development of TI: 
$E_k^{max}={\cal{C}}(\rho_0, T_0, \varepsilon, Q, L)E_{th}(0)$
(see Fig. \ref{fig3}).
The conversion factor ${\cal{C}}$ is a complex function of its variables.
It varies from about 2\% to $\lsim1$\% in our models.
In general, higher $\varepsilon$ and/or $T_0$ values provide higher conversion;
a lower heating level ($Q<1$) supports TI and therefore works in the same 
direction.
Larger boxes, as a rule, also produce more turbulence.
The turbulence is induced on the initial cooling time and decays on a
dynamical timescale, which is typically much longer.
The turbulent Mach number at $t\sim\omega^{-1}_{\rmn{p},0}$ depends 
on the mean temperature at this time.
For nonequilibrium initial conditions ($Q\ll1$ or $Q=0$), this is much lower 
than the initial temperature.
In our fiducial case, ${\cal{M}}_{rms}$ peaks at 8, dropping to 0.4 after 20
initial cooling times.
For equilibrium initial conditions ($Q=1$), turbulent velocities remain
subsonic (${\cal{M}}_{rms}\sim0.3$).
But we would expect higher Mach numbers if the bistable range of pressure were
wider than provided by our adopted cooling function.

The initial gas density determines the number and mass fractions of
thermal phases in the {\em relaxed} state depending on its position 
relative to the valleys and hills on the thermal equilibrium curve.
This is consequence of our choice of constant volume boundary conditions,
which means that the mean density in the box remains constant. 
After the rapid cooling stage, our models with low initial densities 
$\rho_0=1$-$5\times10^{-25}$~g~cm$^{-3}$, high temperature 
$T_0=2\times10^6$~K, 
and $Q\in\{0.3, 1\}$ generate turbulence, evolve through a transient 
three-phase stage, and then relax to a single-phase low-pressure warm ISM.
While turbulence is a generic feature of nonlinear saturation of TI, 
our simulations show that {\em detailed} turbulent properties and the 
nature of 
emerging multiphase medium do depend sensitively on the Mach number and 
effective equation of state controlled by heating and cooling; 
this will be discussed elsewhere.
Two identical simulations, except that cutoffs in initial power spectra were
different ($k_{max}=8$ and 32 on a $128^3$ grid, $L=100$~pc, $Q=0$), 
demonstrated considerable structural differences in density distributions at 
the thermal pancake stage, $t_{tp}$, and surprisingly similar ``chaotic'' 
density structures and identical velocity power spectra at $\sim6\,t_{tp}$, 
when turbulent mixing covered the whole computational domain.
This implies that the imprints of heat conduction in the density power 
spectrum during the linear stage could be erased later by the developing
turbulent cascade.

TI is certainly not the only potential source of turbulence in the ISM, but
it cannot be ignored at least in those scenarios that actively employ TI
to explain the origin and properties of observed objects.
We suggest a paradigm shift concerning the role of thermal instability in the
ISM and the nature of multiphase ISM.
The idea of ``static'' two-phase ISM introduced in late 1960s 
(pressure-confined thermally stable dense clouds embedded in 
rarefied intercloud gas 
forming as a result of TI and subject to phase exchange due to cloud 
evaporation/condensation) must give way to the notion of a dynamic 
multiphase ISM, in which TI induces slowly decaying turbulence and in which 
turbulent diffusion regulates phase exchange processes.
In this new emerging picture, the dense clouds are shapeless random 
aggregations of cold Lagrangian gas parcels; the clouds do not preserve their 
identity in real space on their sound crossing timescale until self-gravity 
tightens the fragments up into a self-gravitating cloud to form stars.
Our results may suggests modifications to the scenario
of a three-phase ISM \citep{mckee.77,mckee90,heiles01a} that are yet to 
be understood.

\acknowledgments

We are grateful to Marco Spaans for providing cooling functions prior to
publication. 
We acknowledge useful conversations with George Field and Chris McKee. 
This work was partially supported by NRAC computer grant MCA98N020N
and utilized the NCSA Silicon Graphics Origin2000 system at the University
of Illinois at Urbana-Champaign.

\clearpage 

\end{document}